\documentclass[acmsmall]{acmart}
\usepackage[linesnumbered,ruled,vlined]{algorithm2e}
\usepackage{multirow}
\usepackage{xspace}
\usepackage{enumitem}
\usepackage{tikz}
\usepackage{cleveref}
\usepackage{tabularx}
\usepackage{adjustbox}
\newcommand{\toolname}{{\sc Athena}\xspace}
\newcommand{\toolnames}{{\sc Athena's}\xspace}

\newcommand{\toolct}{{\sc Athena}$_{ct}$\xspace}
\newcommand{\toolcd}{{\sc Athena}$_{ct+cd}$\xspace}
\newcommand{\toolcmd}{{\sc Athena}$_{ct+cmd}$\xspace}

\newcommand{\bmname}{{\sc Alexandria}\xspace}

\newcommand{\imrepos}{$25$\xspace}
\newcommand{\imcommits}{$910$\xspace}

\definecolor{mycolor}{RGB}{204,204,204}

\newcommand*\circled[1]{\tikz[baseline=(char.base)]{\small{\textbf{
			\node[shape=circle,fill=mycolor,draw=black, inner sep=0.75pt] (char) {\textcolor{black}{#1}};}}}}

\newcommand{\ie}{\textit{i.e.,}\xspace}
\newcommand{\eg}{\textit{e.g.,}\xspace}

\newcommand{\etal}{\textit{et al.}\xspace}

\setcopyright{rightsretained}
\acmDOI{10.1145/3643770}
\acmYear{2024}
\copyrightyear{2024}
\acmSubmissionID{fse24main-p944-p}
\acmJournal{PACMSE}
\acmVolume{1}
\acmNumber{FSE}
\acmArticle{44}
\acmMonth{7}

\begin{document}

\title[Enhancing Code Understanding for IA by Combining Transformers and Program Dependence Graphs]{Enhancing Code Understanding for Impact Analysis by Combining Transformers and Program Dependence Graphs}


\author{Yanfu Yan}
\orcid{0009-0008-2475-6802}
\affiliation{%
  \institution{William \& Mary}
  \city{Williamsburg}
  \country{USA}
}
\email{yyan09@wm.edu}

\author{Nathan Cooper}
\orcid{0000-0003-2498-705X}
\affiliation{%
  \institution{William \& Mary}
  \city{Williamsburg}
  \country{USA}
}
\email{nacooper01@wm.edu}

\author{Kevin Moran}
\orcid{0000-0001-9683-5616}
\affiliation{%
  \institution{University of Central Florida}
  \city{Orlando}
  \country{USA}
}
\email{kpmoran@ucf.edu}

\author{Gabriele Bavota}
\orcid{0000-0002-2216-3148}
\affiliation{%
  \institution{USI Lugano}
  \city{Lugano}
  \country{Switzerland}
}
\email{gabriele.bavota@usi.ch}

\author{Denys Poshyvanyk}
\orcid{0000-0002-5626-7586}
\affiliation{%
  \institution{William \& Mary}
  \city{Williamsburg}
  \country{USA}
}
\email{denys@cs.wm.edu}

\author{Steve Rich}
\orcid{0009-0009-9311-9919}
\affiliation{%
  \institution{Cisco Systems}
  \city{Maryville}
  \country{USA}
}
\email{srich@cisco.com}

\renewcommand{\shortauthors}{Yanfu Yan, Nathan Cooper, Kevin Moran, Gabriele Bavota, Denys Poshyvanyk, Steve Rich}

\begin{abstract}
Impact analysis (IA) is a critical software maintenance task that identifies the effects of a given set of code changes on a larger software project with the intention of avoiding potential adverse effects. IA is a cognitively challenging task that involves reasoning about the abstract relationships between various code constructs. Given its difficulty, researchers have worked to automate IA with approaches that primarily use coupling metrics as a measure of the ``connectedness'' of different parts of a software project. Many of these coupling metrics rely on static, dynamic, or evolutionary information and are based on heuristics that tend to be brittle, require expensive execution analysis, or large histories of co-changes to accurately estimate impact sets.

In this paper, we introduce a novel IA approach, called \toolname, that combines a software system's dependence graph information with a conceptual coupling approach that uses advances in deep representation learning for code without the need for change histories and execution information. Previous IA benchmarks are small, containing fewer than ten software projects, and suffer from tangled commits, making it difficult to measure accurate results. Therefore, we constructed a large-scale IA benchmark, called \bmname, from 25 open-source software projects, that utilizes fine-grained commit information from bug fixes. On this new benchmark, our best-performing approach configuration achieves mRR, mAP, and HIT@10 scores of 60.32\%, 35.19\%, and 81.48\%, respectively. Through various ablations and qualitative analyses, we show that \toolnames novel combination of program dependence graphs and conceptual coupling information leads it to outperform a simpler baseline by 10.34\%, 9.55\%, and 11.68\% with statistical significance.
\end{abstract}



\received{2023-09-29}
\received[accepted]{2024-01-23}

\maketitle

\section{Introduction}\label{sec:intro}

Modern software systems are long-lived, with extensive development and maintenance histories. Many projects experience churn in the developers or teams working on them, and can consist of millions of lines of code~\cite{Shin:TSE11}. As such, understanding the potential cascading impacts of seemingly simple code changes can be a difficult proposition. This comprehension task forms the premise of \textit{impact analysis (IA)} in which a given code change may result in \textit{undesirable side effects}, such as a fault that leads to an erroneous program state, caused by unintended interactions between the changes and other parts of a software system~\cite{kagdi2013integrating, liu2018code}. Thus, the task of IA involves estimating an impact set of entities, usually classes or methods of a software system, from a given change to an entity, also usually a class or a method~\cite{Bohner&Arnold1996b} in the hopes of preventing unintended changes. This process can be cognitively challenging for developers, as reasoning about complex interactions of a software system requires careful comprehension of large volumes of code. Given that many important engineering and maintenance tasks -- such as bug fixing and refactoring -- require code change comprehension, they necessarily require IA as well. This process is typically performed \textit{manually} by developers, but given its complexity, researchers have proposed a range of approaches for automating it.

Past techniques for automated IA have explored using four major types of information: (i) \textit{structural information} (\ie from program dependence graphs), (ii) \textit{semantic or conceptual information} (\ie code similarity), (iii) \textit{evolutionary information} (\ie commit histories), and (iv) \textit{execution information}. Conventional automatic IA techniques~\cite{breech2006integrating, badri2005supporting} have focused on analyzing structural dependencies (\eg control flow dependence) between different code entities to predict change impacts, but they tend to generate large impact sets with lower precision~\cite{li2013survey}. As a result, other IA techniques have chosen to leverage additional information gathered via mining change histories from software repositories~\cite{gethers:icse12, canfora2010using} or program executions~\cite{liu2018code} to generate more accurate impact sets. However, these techniques rely on certain assumptions (\eg sufficient historical data, comprehensive execution profiles), require brittle heuristics, or significantly increase the computational overhead -- making them less practical. These techniques may also ignore the conceptual/semantic information that naturally occurs in code (\eg identifiers) and is key in expressing the underlying intent of code entities. Given that code entities with similar intent likely contribute to similar problem domains, there is another set of IA techniques (\ie conceptual or semantic IA)~\cite{gethers:icse12, kagdi2013integrating, wang2018integrated} which extract vectorized code semantics and compute a similarity-based ranked list of code entities that are potentially impacted by a change. Existing conceptual techniques formulate IA as an information retrieval (IR) task, and typically apply IR-based (\eg latent semantic indexing (LSI)) or machine learning-based (\eg doc2vec~\cite{le2014distributed}) approaches to obtain code representations that capture the semantic relationships between code entities.

The possibility of combining \textit{semantic} and \textit{structural} information specifically for the task of impact analysis has not been well explored~\cite{gyori2017refining}. Such a combination could prove beneficial due to the orthogonal nature of these information sources, and the practicality of forgoing the collection and sanitation of evolutionary or execution information. For instance, semantic coupling can help to relate methods or classes that share similar semantic purposes and hence may impact one another, whereas structural information can help deduce logical relationships between code entities which may appear to be unrelated based upon modeled semantics.

While there is promise in combining semantic and structural information for IA, there is also an opportunity to leverage recent advances in robust semantic models of code. Transformer-based~\cite{vaswani2017attention} neural architectures~\cite{feng2020codebert, wang2021syncobert, wang2021codet5, guo2022unixcoder} have achieved great success in learning rich representations for a variety of code understanding and generation tasks, \eg code search, clone detection, and program repair. These models are typically first pre-trained on large-scale datasets containing unimodal (code-only) and/or bimodal (comment, code) data to learn \textit{generalized} code representations. The models are then fine-tuned on task-specific datasets for downstream code-related tasks. However, despite their demonstrated benefits, none of these models have been applied to IA.

However, adapting transformer-based models of code to the task of IA, and integrating these models with structural information presents at least two major challenges. First, we currently lack large-scale vetted datasets that would allow a neural model to be fine-tuned on \textit{IA-specific} code representations. This is due to the fact that deriving an IA dataset is labor-intensive, as impact sets cannot be easily mined from software repositories without manual validation. Second, while the general code representations produced by pre-trained models could be directly used for similarity calculation for conceptual IA, they still ignore the global context the code finds itself in, \ie the structural dependencies that illustrate how the code is used within a software system. Unlike other code understanding tasks (\ie code search) that can rely solely on isolated code snippets to extract semantics, structural dependencies between code entities also play an important role in IA since the mutually dependent entities are likely to be impacted by each other.

To overcome these limitations, and advance the task of automated IA, we introduce \toolname, which enhances code understanding with Transformer-based neural models~\cite{vaswani2017attention} and structural dependence graphs for capturing relationships among code entities. We perform IA at method-level granularity for code entities in the Java programming language (PL). Specifically, \toolname begins by constructing a software system's dependence graph, where nodes represent methods and edges represent the dependence relationship (\ie call dependence and class member dependence) between methods. We then leverage neural code models including CodeBERT~\cite{feng2020codebert}, UniXcoder~\cite{guo2022unixcoder}, and GraphCodeBERT~\cite{guo2020graphcodebert}, prominent Transformer-based code models, for initial method embedding extraction. These pre-trained neural code models are fine-tuned on a code understanding task, namely code search, to learn richer representations that are aware of the underlying code intent and to potentially transfer the additional knowledge learned from code search to IA. To integrate the global dependence information into local code semantics, the initial method embeddings are further enhanced using an embedding propagation strategy inspired by graph convolutional networks (GCN)~\cite{kipf2016semi} based on the constructed dependence graphs.

Evaluating our proposed approach effectively also presents challenges. Existing IA benchmarks tend to be outdated and are constructed from original/unvetted commits, but as highlighted in multiple prior studies~\cite{kochhar2014potential, kirinuki2016splitting, wang2019cora, mills2020relationship}, \textit{tangling} has a high prevalence in these commits which is likely to affect the reliability of evaluation results of previous IA techniques on these benchmarks. Therefore, to evaluate \toolname for the task of IA, we created a large-scale IA benchmark, called \bmname, that leverages an existing dataset of fine-grained, manually untangled commit information from bug-fixes~\cite{herbold2022fine}. The benchmark consists of \imcommits commits across \imrepos open-source Java projects, which we use to construct 4,405 IA tasks -- where each task consists of a query method and a set of impacted methods. Using the standard information retrieval metrics of mRR, mAP, and HIT@10, we find that \toolname significantly (based on statistical tests) improves over the best-performing conceptual IA baseline by 10.34\%, 9.55\%, and 11.68\%, respectively. In aggregate, we make the following contributions:

\begin{itemize}[leftmargin=1.2em]

    \item A new large-scale evaluation benchmark for impact analysis, called \bmname, composed of 4,405 IA tasks from \imcommits commits of \imrepos open-source software systems;

    \item The first application of Transformer-based neural models to impact analysis for semantically rich code representations;

   \item \toolname, a novel approach that first integrates global dependence information into local code semantics to advance automated impact analysis;

   \item A comprehensive empirical evaluation that demonstrates that \toolname achieves state-of-the-art improvements compared to the conceptual IA baseline;

    \item A thorough set of ablations showing that the improvements are attributable to the application of the Transformer-based neural model and the integration of structural dependence information;

    \item A comprehensive online appendix~\cite{athena-tool} and archived replication package~\cite{athena-replication} that contain the code for \toolname, our IA benchmark \bmname, and our experimental infrastructure to allow for replication.

\end{itemize}
\section{Background \& Related Work}\label{sec:back}

\subsection{IA Techniques}

Typical IA techniques require a seed/starting entity to perform the analysis. Some start with a change request~\cite{gethers:icse12, torchiano2010impact} in natural language form, while most start with code entities~\cite{poshyvanyk2009using, kagdi2013integrating, liu2018code} at different levels of granularity (\eg classes, methods, statements) since developers can usually identify at least one code entity that needs to be changed by using feature location techniques~\cite{dit2013integrating} and their software development knowledge. The output of the IA (\ie estimated impact set) is usually at the same granularity-level as the seed entity. Given that the class/file-level IA~\cite{torchiano2010impact} is too coarse and the statement-level IA~\cite{gyori2017refining} is too costly, most existing techniques choose to conduct IA at the method level~\cite{wang2018integrated, liu2018code}. Moreover, Java, as one of the most commonly used object-oriented programming languages (PLs), has been selected as the primary focus of IA more often than any other PL (\eg C~\cite{gyori2017refining}).

In general, IA comprises two branches of techniques. One is to predict/infer \textit{potential} impact of all possible changes~\cite{cai2015comprehensive, cai2016distia, gyori2017refining} (\ie dependence analysis); the other is to reason about the \textit{actual} impact sets of code changes~\cite{kagdi2013integrating, liu2018code, wang2018integrated}. Specifically, the first branch assesses the user-perceived accuracy by creating the ground-truth impact set based on static program dependence analysis or dynamic execution differencing, since it regards the real ground truth as unknown. However, identifying the full set of dependencies based on static analysis is uncertain, and execution differencing relies on certain test cases and executions, which cannot cover all possible dependencies either. Cai~\cite{cai2020reflection} gives a comprehensive summary of the first branch of techniques, while our approach falls into the second category, and we will now introduce the related techniques within this category in detail.

Existing IA techniques in the second category can be further divided into four types based upon the information they analyze, \ie structural, conceptual/textual, evolutionary, or dynamic. Conventional IA approaches~\cite{badri2005supporting, breech2006integrating} that use program graphs or slicing tend to generate very large impact sets~\cite{li2013survey}, and most importantly, they ignore the conceptual information encoded in the code (\eg identifiers) which is also important for expressing the intent of code entities. Since code entities with similar intents likely contribute to similar problem/solution domains, conceptual IA techniques~\cite{poshyvanyk2009using, torchiano2010impact, kagdi2013integrating, wang2018integrated} typically apply IR-based (\eg LSI) or machine learning-based (\eg doc2vec~\cite{le2014distributed}) approaches on code to extract vectorized code semantics and estimate impact sets by computing a cosine similarity-based ranked list of code entities. Poshyvanyk \etal~\cite{poshyvanyk2009using} quantitatively show that conceptual coupling is superior to structural coupling-based measures for IA. Moreover, some IA techniques analyze evolutionary couplings~\cite{zimmermann2005mining, sherriff2008empirical, jashki2008towards} mined from multiple historical releases/commits of version control systems in order to discover frequent co-change patterns to predict current change impacts, but sufficient historical data is not always available (\eg for new projects), and sometimes previous change patterns may be outdated and misleading. In addition, dynamic IA~\cite{breech2004online, liu2018code} utilizes execution information (\eg execution traces, relations) to compute more accurate impact sets, but the computational overhead is much greater than that of static IA. The quality of dynamic techniques relies heavily on the representativeness of the test suites and/or profiles gathered during program execution. Industrial case studies~\cite{borg2016supporting, tao2012software, acharya2011practical, de2016industrial, gyori2017refining} indicate a preference for static IA techniques over dynamic ones, as there is a lack of published studies reporting the adoption of dynamic IA~\cite{cai2020reflection}.

To further improve the accuracy of impact set estimation, some research attempts to combine existing techniques. Kagdi \etal~\cite{kagdi2010blending} blend conceptual and evolutionary analysis, showing additional advantages over using either of them alone. Gethers \etal~\cite{gethers:icse12} further augment them with dynamic analysis to obtain more accurate impact sets. It is worth noting that these two hybrid techniques are only compared with their own variants (\ie using only one of the components) to validate their effectiveness. A recent work~\cite{liu2018code} combines dynamic analysis with structural analysis (\ie data and call dependencies), demonstrating that dynamic data-sharing dependencies are complementary to dynamic call dependencies.

Our approach belongs to the set of hybrid analysis-based IA techniques as \toolname extracts code semantics and dependencies and computes a ranked list for impact set estimation. Therefore, it avoids the associated limitations and drawbacks of other categories of techniques (\ie evolutionary and dynamic analysis) while retaining the benefits of multiple information sources. LSI is the most frequently used model to obtain code semantics for conceptual IA~\cite{poshyvanyk2009using, kagdi2010blending, gethers:icse12}. The latest and most closely related work to ours is that of Wang \etal~\cite{wang2018integrated}, which integrates LSI with doc2vec to enhance code semantics by considering the context of each code token within the code entity. They quantitatively show that the combined model outperforms using LSI alone on IA.

Different from existing conceptual IA techniques, our approach (i) leverages advanced Transformer-based code models to obtain more meaningful code representations, and (ii) further enhances code semantics via embedding propagation based on structural dependence graphs. To the best of our knowledge, our approach is the first IA technique that integrates global structural information into local code semantics based on only a single release of the source code, without any additional information (\eg previous releases and/or execution information). Given that Wang \etal~\cite{wang2018integrated} have not made their implementation publicly available, we directly use LSI and doc2vec independently as conceptual IA baselines for our work. This also allows us to compare the performance of different models for code semantics extraction when they are individually applied for IA.

\subsection{IA Benchmarks}

Existing IA benchmarks~\cite{liu2018code, cai2015comprehensive, gethers:icse12} are typically constructed in two ways. The first type of construction considers ground-truth impact sets to be unknown and tries to create them using program dependence analysis~\cite{cai2014diver, cai2015prioritizing} or execution differencing~\cite{cai2015comprehensive, cai2016distia, cai2016diapro, gyori2017refining}. However, computing a full set of program dependencies~\cite{cai2020reflection} is an undecidable problem. As such they are usually generated based on artificial changes and/or by sampling changes in real open-source projects. All possible changes to a code entity (only involving \textit{one} certain release of code repository) are used as the seeding entities.

The other more popular way for constructing IA benchmarks involves building multiple co-changed sets of code entities, each of which are collected based on \textit{two} consecutive commits~\cite{liu2018code} or several grouped commits~\cite{wang2018integrated}. All entities within a co-changed set are assumed to be impacted by each other. To construct the ground truth, one~\cite{kagdi2013integrating} or a few code entities~\cite{liu2018code} in the co-changed set are selected as the seed entity, and the remaining ones serve as the real impact set. Existing benchmarks/case studies in this category usually consist of 3--6 open-source repositories, and the commits used are either bug fixing commits only~\cite{jiang2020automatic} or dominated by bug fixing commits~\cite{gethers:icse12}. However, the prevalence of \textit{tangling}~\cite{herzig2013impact, kirinuki2014hey, mills2020relationship, herbold2022fine} existing in commits negatively affects the reliability of evaluation performance of techniques (\eg bug localization~\cite{mills2020relationship}, defect detection~\cite{herzig2016impact}) that rely on commit data for testing due to the presence of noise. Tangled commits refer to the changes to software which address multiple concerns at once. For example, a (original) commit which claims to be fixing a bug, may not only fix the bug but also include additional unrelated changes (\eg refactorings).
While we have limited knowledge of the exact impact of tangled commits on the reliability of IA technique evaluations, the potential for impact is clear --- in tangled commits the co-changed code entities within a commit do not all contribute to a single concern (\eg bug fixing) and thus are not necessarily impacted by each other, leading to inaccurate ground-truth impact sets. Given that prior studies have confirmed the prevalence of tangled commits~\cite{herbold2022fine}, it is highly likely that evaluations of past techniques were affected by this phenomenon.

Our \bmname dataset falls into the second category of IA benchmark, but with a \textit{notable key difference} --- it is built from untangled bug fixing commits~\cite{herbold2022fine}. Herbold \etal's work quantitatively shows that tangled commits have a high prevalence, and the authors manually untangle them by annotating line-level change types. By utilizing only the co-changed code entities that have been manually verified to contribute to one concern (\ie bug fix), our benchmark contains more reliable ground-truth impact sets, and this favorable characteristic is demonstrated quantitatively through experiments. To the best of our knowledge, our \bmname is the first IA benchmark whose ground-truth impact sets are built from manually-validated untangled commits. Moreover, \bmname contains \imcommits commits from \imrepos systems, making it larger than past benchmarks.

\subsection{Code Representation Learning}

Traditional IR approaches (\eg LSI, Term Frequency - Inverse Document Frequency (TF-IDF), Latent Dirichlet Allocation (LDA)) were first used to generate vectorized code representations in order to support SE tasks. They typically require building a corpus from all documents (code artifacts) and then represent code by measuring the importance of each code token to a document in the corpus and/or exploiting co-occurrences of code tokens based on singular value decomposition (SVD) or Bayesian topic modeling. However, these IR approaches treat the code as \textit{bag-of-words}, ignoring the order and semantics of code tokens. Thus, neural networks have been employed to obtain more meaningful code representations. For instance, word2vec~\cite{mikolov2013efficient} takes into account each individual token and its context tokens by using a sliding context window during training. Furthermore, doc2vec~\cite{le2014distributed} could learn a paragraph vector for the code of variable length, instead of using an average representation of the code tokens as word2vec does. Subsequently, more and more end-to-end deep models (\eg Bi-RNN~\cite{cho2014properties}, TextCNN~\cite{kim2014convolutional}, Self-Attention~\cite{vaswani2017attention}) have been used to extract code embeddings.

Recently, the fine-tuning-after-pre-training scheme~\cite{devlin2018bert, yang2019xlnet, raffel2020exploring, brown2020language} has achieved great success in NLP tasks wherein a model is pre-trained on large-scale text in a self-supervised manner to learn general representations, and then fine-tuned for specific downstream tasks on a more limited dataset. The Transformer~\cite{vaswani2017attention} architecture stands out as the most representative encoder backbone for this scheme. With the advent of large-scale code datasets (\ie CodeSearchNet~\cite{husain2019codesearchnet}), this scheme has also been increasingly applied to learn code representations and automate software engineering tasks~\cite{wang2021codet5, guo2022unixcoder, wang2021syncobert, ahmad2021unified}. CodeBERT~\cite{feng2020codebert} was one of the first Transformer-based NL-PL pre-trained models for supporting various code-related tasks. It distinguishes between the PL and the NL modality and captures their semantic connection during pre-training. However, it only utilizes the sequential information of the bi-modal data while ignoring the inherent structure of code. Therefore, GraphCodeBERT~\cite{guo2020graphcodebert} further incorporates data flow information within methods into sequenced code snippets during pre-training, resulting in enhanced code embeddings. Other pre-trained models, such as UniXcoder~\cite{guo2022unixcoder}, encode abstract syntax tree (AST) information to produce syntax-aware code embeddings. We explore the use of pre-trained CodeBERT, GraphCodeBERT, and UniXcoder representations for \toolname, which are then fine-tuned on the code search task to extract initial code embeddings for performing IA. However, any Transformer-based code model can serve as the encoder backbone of our approach.
\section{\toolname}\label{sec:approach}

\begin{figure*}[t]
    \centering
    \includegraphics[width=\linewidth]{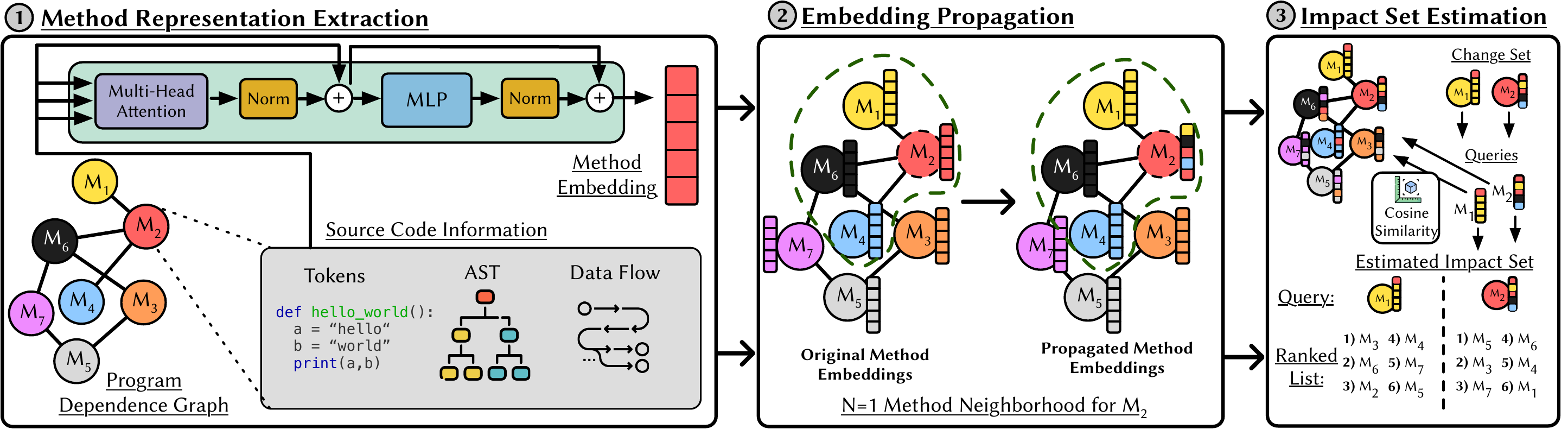}
    \caption{Overview of the workflow of the \toolname impact analysis approach.}
    \Description{A three-stage pipeline. Stage one extracts an initial embedding for
    every method of a software system with a fine-tuned Transformer-based code model.
    Stage two propagates embeddings between neighboring methods over a dependence graph
    whose edges encode call and class member dependencies. Stage three ranks all methods
    in the corpus by the cosine similarity of their augmented embeddings to the query
    method, producing the estimated impact set.}
    \label{fig:athena}
\end{figure*}

In line with previous conceptual IA techniques~\cite{gethers:icse12, kagdi2013integrating, wang2018integrated}, we formulate impact analysis as an information retrieval task where if a developer intends to modify a method (\ie query/seed method) in a software system, \toolname will return a ranked list of other methods being potentially impacted in descending order of likelihood. All methods but the query are used as the search corpus. Formally, for a software system $S$ containing a set of methods $S = \{ m_1, m_2, ..., m_n \}$, a potential change to one of the methods $m_i \in S$ triggers \toolname to rank all other methods thus estimating the impact set.

\Cref{fig:athena} provides an overview of \toolname. \toolname begins by building a dependence graph among all methods across an entire software system, where nodes represent methods and edges represent dependence relationships between methods. Each method is processed by a state-of-the-art Transformer-based code model (\eg GraphCodeBERT) to obtain an initial method representation by considering the context that exists within the method. These neural code models are then fine-tuned on the code search task to generate richer code representations and potentially transfer the additional knowledge learned from code search to IA. Next, \toolname analyzes the global dependencies and propagates information from the ``neighbor'' method nodes in the dependence graph to a given target method. Specifically, each initial method embedding is updated/augmented based on a propagation strategy inspired by Graph Convolutional Networks (GCNs)~\cite{kipf2016semi} so that the information of global dependences is integrated into its local code semantics. To obtain a final ranked list, the cosine similarity between the augmented representations of a given query method and each method in the corpus is computed. We next discuss each step of \toolname in detail.

\subsection{Dependence Graph Generator}

The initial step of \toolname is to build a static dependence graph generator to capture method dependencies across a software system. Essentially, we identify two methods as having dependencies if there exists a caller-callee relationship between them (\ie call dependence) and/or if they belong to the same class (\ie class member dependence). While certain existing tools like WALA~\cite{fink2012wala} and Soot~\cite{Soot} can produce static call graphs for Java, they require JVM bytecode as input, thus necessitating compilable source code. Although the latest version of Soot provides source code analysis, it limits the source code to Java 7 and still requires internal compilation. These tools thus increase preprocessing time for IA and negatively affect scalability. To better integrate the graph generator into \toolname and capture both call and class member dependencies, we developed our own tool to generate static dependence graphs, which simply takes the source code of a software system as its input.

A dependence graph can be formally defined as $G = (V, E)$, where $V$ denotes a set of method nodes and $E$ denotes a set of edges representing the method dependence relationships. Since impact analysis is usually performed on the production entities (\ie excluding the entities for testing)~\cite{liu2018code}, we first collect all \texttt{\small .java} production source files in a software system and use the Tree-Sitter~\cite{tree_sitter} library to identify all methods contained in these files. The library enables constructing a specific syntax tree for each file and supports searching for various patterns (\eg method calls, method declarations) in the tree. All identified methods then serve as the nodes of the dependence graph. To precisely locate each method and facilitate the process of method representation extraction, we attach to each method node the complete method content (\ie the method declaration with its body), the name of the class it belongs to, and the package path.

Next, we construct the edges for the dependence graph. To capture the class member dependencies, the edges are added between each pair of the methods in the same class. As for the call dependencies, we utilize the Tree-Sitter library to identify all the method invocation statements (\eg \texttt{receiver.method()}) within each method and resolve these statements by finding its callee methods. The edges are then added between each pair of caller-callee methods. In general, we traverse upwards from each invocation statement to find where the \texttt{receiver} is introduced by analyzing the declaration statements and the arguments of the caller method. It is then easy to obtain the class name of the callee method and the package path it belongs to. In order to locate the callee method based on the class name and the package path, we utilize both the method name and the number of arguments (rather than the complete signature) to ensure the efficiency and scalability of our generator. When the callee method is overloaded with the same number of arguments, we add edges from the caller method to each of these overloaded callee methods. It is worth noting that combining the method name and the number of arguments filters out considerably more overloaded methods than using the method name alone.

Although we can add directed edges from caller to callee methods, their semantics are actually interrelated and mutually affect each other when performing IA. Thus, by using our tool, the dependence graph is constructed in an undirected manner. Moreover, edges representing class member dependencies are distinguished from those representing call dependencies by attaching each edge to its property (\ie call or class member dependence). If two methods have both types of dependencies, we add two edges with different properties between them.

\subsection{Code Representation Extraction}

We then use one of three Transformer-based code models (CodeBERT, UniXcoder, or GraphCodeBERT) to extract initial method embeddings for performing IA, as shown in \Cref{fig:athena}-\circled{1}. In the case of GraphCodeBERT, it goes beyond sequential code information by considering the inherent structure of code (\ie data flow) to encode the ``where-the-value-comes-from'' relation between variables.
In this model, the input is encoded by a multi-layer bidirectional Transformer containing a sequence of self-attention and feed-forward layers (\ie a multi-layer perceptron (MLP)) with normalizations.

These pre-trained models can directly produce code embeddings, but the self-supervised objectives used during pre-training are quite different from IA, and most importantly, the representations are not specifically learned for Java but generally for multiple PLs. Although these neural models can be further fine-tuned for downstream tasks, neither GraphCodeBERT nor other Transformer-based code models have been fine-tuned or evaluated for IA, due to the absence of large available IA training/fine-tuning datasets. IA belongs to a general family of code understanding tasks (and hence is not generative), and there are two other downstream understanding tasks that have been extensively researched and evaluated --- namely code search and clone detection. Code search aims to retrieve relevant code given a NL query, while clone detection aims to predict whether two code snippets can output similar results when given the same input. We leverage code search as a proxy to potentially transfer additional knowledge learned from code search during fine-tuning to enhance code semantics for IA. Although clone detection may initially seem more closely aligned with IA, we do not use it because (i) datasets such as BigCloneBench~\cite{lu2021codexglue, svajlenko2014towards} that could be used for fine-tuning do not include comments, which are likely to enhance code understanding; and (ii) instead of generating separate code embeddings, the fine-tuned neural model for clone detection concatenates two code snippets as a whole and generates only one embedding for them, thus making the subsequent embedding propagation process more difficult. They typically add a classifier on top of the Transformer-based encoder to directly produce the probability of whether two code snippets can yield similar results.

To fine-tune our neural code models for code search, we follow the pipelines recommended in their corresponding papers. For example, for GraphCodeBERT, our best-performing model, we follow the authors' recommendation~\cite{guo2020graphcodebert} to use a Siamese framework on the CodeSearchNet~\cite{husain2019codesearchnet} Java split dataset. CodeSearchNet consists of 2.3 million functions in six programming languages paired with NL descriptions (\ie comments). The CodeSearchNet Java split has been filtered with handcrafted rules by Guo \etal~\cite{guo2020graphcodebert} to remove low-quality data, and contains 164,923 bimodal (comment, code) pairs. Each code snippet in the paired data is a method from a GitHub repository with all comments removed, and the corresponding comment is extracted from the first line of the method's documentation comment. The objective of fine-tuning is to map the code and its comment onto vectors that are close to each other in order to learn high-level intent-aware code semantics. During fine-tuning, the comment and code (with data flow extracted) are separately fed into a comment encoder and a code encoder. These two encoders have identical model architectures (\ie GraphCodeBERT) and are initialized from the pre-trained GraphCodeBERT parameters (\ie weights and biases). The parameter updating is synchronized across both encoders during fine-tuning based on the standard cross entropy loss. We use the AdamW~\cite{loshchilov2019decoupled} optimizer and the same hyperparameters (\eg number of epochs, learning rate, and batch size) recommended by Guo \etal~\cite{guo2020graphcodebert} for parameter updating, and the whole process was performed on an Ubuntu 20.04 server with an NVIDIA A100 40GB GPU. The fine-tuned GraphCodeBERT is expected to generate more meaningful representations of code that are aware of the underlying intent.

When performing IA, we need to first preprocess the method content attached to each method node in the generated dependence graph. Taking GraphCodeBERT as an example, we first follow the preprocessing procedure of CodeSearchNet~\cite{husain2019codesearchnet} by extracting the initial line of the documentation comment and the code-only data. The code is further parsed into an abstract syntax tree (AST), the leaves of which are used to identify the variable sequence for the data flow construction. The input to the fine-tuned GraphCodeBERT for IA is the concatenation of the comment, the source code, and the set of variables, \ie $X = ([CLS], A, [SEP], C, [SEP], V)$ or $X = ([CLS], C, [SEP], V)$. $A$, $C$, and $V$ stand for the comment token sequence, code token sequence, and variable sequence, respectively. $[CLS]$ is a token for learning aggregated information from the entire sequence during training, and its final representation is typically used for classification-related tasks. $[SEP]$ is a separation token used to split two data types. Edges are added between variables in the variable sequence where a data flow relationship exists, and the variables are aligned across source code and data flow. The input is then processed by the fine-tuned encoder, and we take the average output of all the hidden states of the last layer as the method representation. The input sequence length is set to 256 and the output representation dimension is 768 to maintain consistency with GraphCodeBERT. Finally, the initial method embeddings are generated for all method nodes in the dependence graph of a given software system.

\subsection{Embedding Propagation}

While the initial embeddings effectively capture meaningful code semantics via the self-attention mechanism, they are limited to local context and lack the global dependence information of methods. To further improve code understanding, we utilize an embedding propagation strategy that updates each method embedding by propagating the embeddings of its neighbor methods based on the constructed dependence graph $G$, thus integrating the information of global structural dependence into local code semantics. We visualize this process in \Cref{fig:athena}-\circled{2}. Formally, this is represented as $m'_i = f(m_i, m^{nebr}_1, m^{nebr}_2, ..., m^{nebr}_k),$ where $m_i$ is the method being updated through the embedding propagation strategy $f$ with its neighbors $m^{nebr}_j (1 \le j \le k)$. In particular, our embedding propagation strategy is inspired by the Graph Convolutional Network~\cite{kipf2016semi}, which adopts layer-wise propagation on neural networks motivated by a localized first-order approximation of spectral graph convolutions:

\begin{equation}
    M' = \sigma( \tilde{D} ^ {-\frac{1}{2}} \tilde{A} \tilde{D} ^ {-\frac{1}{2}} M W),
\end{equation}

\noindent where $\sigma$ represents an activation function and $W$ is a trainable weight matrix. $\tilde{A} = A + I_N$ denotes the adjacency matrix of a graph $G$ with self-connections. $I_N$ is the identity matrix and $\tilde{D}_{ii} = \sum_j \tilde{A}_{ij}$. This propagation strategy has been modified using a renormalization method~\cite{kipf2016semi} in order to mitigate the effects of numerical instabilities and exploding/vanishing gradients when matrix multiplication operators are repeated during the training of the deep neural network. Since we do not train our dependence graph $G$ in this phase, our embedding propagation strategy is directly derived from the first-order approximation of localized spectral filters on graphs~\cite{hammond2011wavelets, defferrard2016convolutional}, which can be summarized as follows:

\begin{equation}
    M' = (I_N + w D ^ {-\frac{1}{2}} (A^c+A^{cm}) D ^ {-\frac{1}{2}}) M.
\end{equation}

\noindent $M \in \mathbb{R}^{N \times F}$ represents the matrix of all method embeddings with respect to the dependence graph $G$ and $M' \in \mathbb{R}^{N \times F}$ stands for the matrix in which each method embedding is updated by its neighbor method embeddings. $N$ denotes the number of method nodes and $F$ denotes the dimension of each method embedding (\ie $768$). $A^c$ is the adjacency matrix based on call dependence edges of $G$, while $A^{cm}$ is the one based on class dependence edges. Neither of them contains self-connections. $D$ is the degree matrix of $(A^c+A^{cm})$ for normalization with respect to both rows and columns. $w$ is a constant that is responsible for balancing the information between a method and its neighbor methods. According to this formula, if a method exhibits both call and class member dependencies with a neighbor method, the embedding of this neighbor method will be propagated/aggregated twice to the target method embedding. Intuitively, methods sharing multiple dependencies are inherently more closely related than those with just a single type of dependency. Moreover, in order to evaluate the effect of the distance of neighbor methods used for embedding propagation, neighbor methods at other orders (hops) are also utilized in addition to the direct neighbors:

\begin{equation}
    M' = (I_N + w \sum_i D_i ^ {-\frac{1}{2}} (A^c_i + A^{cm}_i) D_i ^ {-\frac{1}{2}}) M,
\end{equation}

\noindent where $1 \le i \le 3$, since we take into account at most the neighbor methods within three orders due to computational constraints. After the embedding propagation strategy has completed, all of the identified methods in a given software system will have an augmented embedding calculated by propagating the \textit{original} method embedding from neighbors to the target method, as illustrated at the top of \Cref{fig:athena}-\circled{3}.

\subsection{Impact Set Estimation}

Finally, as illustrated in \Cref{fig:athena}-\circled{3}, \toolname computes the cosine similarity between the augmented embedding of a given query method and the augmented embeddings of each of the methods in the search corpus. Based on the cosine similarity scores, \toolname returns a ranked list in descending order to help developers find other methods that are possibly affected and likely to be modified.
\section{Experimental Design}\label{sec:design}
To evaluate \toolnames effectiveness in IA, we investigate the following research questions (RQs):

\begin{enumerate}[label=\textbf{RQ$_\arabic*$:}, ref=\textbf{RQ$_\arabic*$}, wide, labelindent=5pt]\setlength{\itemsep}{0.2em}
    \item \label{rq:athena}{\textit{How effective is \toolname with/without embedding propagation when compared with conceptual baselines on the task of impact analysis? }}
    \item \label{rq:dependence}{\textit{How do call and class member dependencies improve \toolnames effectiveness in IA? }}
    \item \label{rq:ablation}{\textit{How well does \toolname perform on IA based on different configurations (\eg using other Transformer-based pre-trained code models)? }}
    \item \label{rq:tangled}{\textit{How does the tangled benchmark affect the reliability of IA evaluation results?}}
    \item \label{rq:impact}{\textit{How do properties of different impact analysis tasks affect our studied techniques?}}
\end{enumerate}

\subsection{Impact Analysis Benchmark: \bmname} \label{sec:benchmark}

Our IA benchmark \bmname is constructed from manually untangled bug fixing commits~\cite{herbold2022fine} in order to generate more reliable ground-truth impact sets. Multiple prior studies~\cite{kirinuki2016splitting, wang2019cora, nguyen2013filtering}, supported by manual validation, have consistently shown that tangled commits naturally occur in codebases. However, all existing IA benchmarks~\cite{kagdi2013integrating, liu2018code, wang2018integrated}, built directly from these original/unvetted commits, inaccurately assume that all co-changed entities in a commit address one single concern and are thus impacted by each other. The unvalidated data (\ie (query, ground-truth impact set) pairs) is likely to be noisy, which can affect the reliability of experimental results of previous IA techniques.

Recently, Herbold \etal~\cite{herbold2022fine} introduced a large dataset covering 3,498 commits from 28 Java projects, with the purpose of studying the tangling that occurs in bug fixing commits. All selected projects are from the Apache Software Foundation and were developed by contributors from the open source community or industry. These projects cover diverse application domains, such as build systems (\eg \textit{ant-ivy}), web applications (\eg \textit{jspwiki}), and general purpose libraries (\eg \textit{commons}). In this dataset, each changed line was annotated with its type of change, whether it was modified to fix a bug, or was a change to tests, whitespace, a documentation change, a refactoring, or unrelated feature improvement. The data were annotated by four participants, and consensus was obtained if at least three participants agreed on the annotation to ensure accuracy.

While some existing datasets~\cite{kirinuki2014hey, kochhar2014potential, mills2020relationship} also manually untangle the commits, they either cover a limited sample of commits or typically perform untangling at the commit or file level, which is relatively coarse-grained so that the validated co-changed entities cannot be identified at method-level. Therefore, we constructed our IA benchmark based on the fine-grained untangled dataset~\cite{herbold2022fine} allowing us to know exactly which methods are changed for addressing one single concern, thereby generating reliable ground-truth for evaluation.

\textbf{Co-Changed Set Construction.} To create evaluation IA tasks, we systematically mined the dataset of Herbold \etal~\cite{herbold2022fine}. By utilizing only the co-changed code entities that have been rigorously manually verified to contribute to one concern, our benchmark \bmname contains more reliable ground-truth impact sets. Specifically, for each changed line in production code files labeled as \textit{``contributes to the bug fix''}, we added the corresponding method to our benchmark by recording the GitHub Diff URL, repository name, commit ID, parent commit ID, file path, method name, and the line numbers indicating where the method starts and ends. Since the dataset~\cite{herbold2022fine} does not provide method-related information, such as method names and the line numbers of method boundaries, we employed the srcML library~\cite{collard2013srcml} to locate each changed method based on the labeled changed line numbers. We utilized the snapshot/release of a software system that corresponds to the parent commit ID, as that is the state in which the change would be applied. Then, for each parent commit, we formulate a co-changed method set based on concurrently changed methods. Since there is no clear indication of a query/seed method, \ie which method would be changed ``first'' in the commit, we treat each method in the co-changed method set as a potential query, whereas the remaining ones constitute the ground-truth impact set. From a developer's point of view, they usually at least know where the change starts and want to know which other methods need to be modified. We further post-process the dataset to exclude commits that contain only one changed method.

\textbf{IA Task Definition and Settings.} Formally, for each co-changed method set $M = \{m_1, m_2, ..., m_n\},$ $n \geq 2$, we perform IA with a query being $ \forall~m_i \in M $ and the corresponding ground-truth impact set being $M - {m_i}$. We consider three different settings wherein the search corpus differs. In the first setting (\textbf{Setting 1 - whole}), the search corpus includes all methods except the query in all production files from the corresponding snapshot of the software system. This setting provides a comprehensive evaluation scenario where all methods in the software system are taken into consideration. The similar process of formulating co-changed methods into IA tasks has been widely adopted by past work to assess IA approaches~\cite{kagdi2013integrating, liu2018code, gethers:icse12}. In practice, conceptual IA techniques will generate a ranked list of methods in the corpus and developers would determine whether a method should be modified by inspecting the corpus in the given order. After analyzing our benchmark, it was observed that methods in the same class are more likely to be changed together. To account for this and mitigate potential biases introduced by IA approaches that equally prioritize methods within the same class as the query, we formulate two more specific task settings. In our second setting, the methods in both the ground-truth impact set and the search corpus are from the same class as the query (\textbf{Setting 2 - inner}).
In our third setting, the methods in both the ground-truth impact set and the search corpus are from different classes than the query (\textbf{Setting 3 - outer}).

\textbf{Dataset Statistics.} Two software projects (\ie \textit{santuario-java} and \textit{wss4j}) in the dataset of Herbold \etal~\cite{herbold2022fine} are no longer accessible, and for the software project \textit{eagle}, we were unable to build any valid co-changed method sets, \ie the size of every co-changed set was less than two. As a result, our benchmark contains \imrepos Java software projects, and the lines of code (LOC), number of commits, and number of tasks for each project are shown in~\Cref{tabs:repo_results}. Moreover, for each of the three settings, \Cref{tab:dataset} shows the number of tasks, the number of commits, and the average number of methods in the ground-truth impact set and in the search corpus, respectively. Compared to Setting 2 (inner), which requires retrieving roughly four or five affected methods out of an average of 30 methods, Setting 3 (outer) is far more challenging, requiring roughly 17 methods to be retrieved from a larger corpus with an average of 3,440 methods.

\begin{table}[t]
\vspace{-1em}
\centering
\small
\caption{Dataset statistics of the \bmname evaluation benchmark. The last two columns report the average number of methods in the ground-truth impact set and in the search corpus.}
\label{tab:dataset}
\begin{tabular}{l|c|c|c|c}
\toprule
Settings  &  \# queries   &  \# commits    & ground-truth set   & corpus \\  \hline
1 - whole    & 4,405  & 910    & 15.14 &   3,346    \\
2 - inner    & 3,379  & 734    & 4.47  &   30    \\
3 - outer    & 2,999  & 444    & 17.21  &   3,440    \\  \hline

\end{tabular}
\vspace{-2em}
\end{table}

\textbf{Tangled Counterpart.} To analyze the effect of tangling commits on the evaluation of IA techniques, we also construct a benchmark without manual untangling, similar to what previous IA benchmarks did~\cite{liu2018code, wang2018integrated, kagdi2013integrating}. Specifically, we directly construct co-changed method sets from original/tangled commits, so the bug fix changes are likely to be tangled with refactoring and unrelated improvement changes. Then, we compare the \bmname dataset with its tangled counterpart in terms of the tasks with inconsistent (query, impact set) pairs. We observe that 606 tasks from 50 commits (Setting 1) in \bmname would have had inaccurate ground-truth impact sets without untangling. Further, the tangled \bmname dataset has 856 tasks (out of 4,655) from 81 commits that are inaccurate with respect to (query, ground-truth impact set) pairs. The increase in the number of tasks and commits is due to an increase in the size of co-changed method sets, \ie more changed methods (for refactoring/unrelated improvement) are used as queries, and some previously filtered commits with co-changed sets smaller than two are added back.

\subsection{Evaluation Metrics}\label{sec:metrics}

We use standard information retrieval metrics to measure the effectiveness of \toolname, namely mRR (mean Reciprocal Rank), mAP (mean Average Precision) and HIT@k. For each task, the ranked list generated by \toolname is compared with the ground-truth impact set. Specifically, we computed the \textit{rank} of the \textit{first} truly affected method found in the ranked list, indicating the number of methods developers need to inspect before finding the first one that requires modification. The \textit{reciprocal rank} is then calculated for each task, and these values are averaged across all tasks to derive the final mRR score. Furthermore, we compute the AP score for each task and average these scores across all tasks to obtain the final mAP score. AP is the average of precision values calculated after \textit{each} method in the ground-truth impact set is retrieved, which approximates the area under the uninterpolated Precision-Recall curve. mAP scores measure the ability of the approach in helping developers identify all possibly affected methods. Moreover, we use HIT@$k$ to measure the proportion of successful tasks for the cut point $k$. A successful task means that the approach has found at least one truly affected method among the top-$k$ results it returns.

Many IA techniques~\cite{liu2018code} rely on \textit{Precision, Recall} and \textit{F-measure} for evaluation since they consider IA as a binary classification task by finding possibly affected methods based on structural/evolutionary/dynamic dependencies. Therefore, what these techniques produce is not a ranked list, but an unordered estimated impact set, which is then directly compared with the ground truth impact set to compute an F-score (\ie the harmonic mean of the \textit {Precision} and \textit {Recall} values). However, conceptual IA techniques~\cite{wang2018integrated, kagdi2013integrating, gethers:icse12} formulate IA as an information retrieval task but still adapt prior Recall/Precision/F-score metrics to the IR context. We argue that IR metrics provide a more realistic representation of the potential benefits that conceptual IA approaches may actually provide to a developer in a recommender system setting. Furthermore, mAP score is more accurate than F-measure because it analyzes \textit{Precision-Recall} relationship globally rather than just based on the mean value calculation.

\subsection{Baselines}
We compare our approach, \toolname, with three baseline approaches that extract code semantics for intent-aware IA. Specifically, two traditional IR-based approaches (\ie TF-IDF and LSI) and a deep learning-based model (\ie doc2vec~\cite{le2014distributed}) are used as our conceptual IA baselines. To use IR for IA, we first build a corpus using all production methods from a specific snapshot/commit of a software system. For each code token in a method, we calculate its term frequency (TF), which represents the number of times the token appears in the method, and its inverse document frequency (IDF), which is based on the number of occurrences of the code token across all methods in the corpus. Each method in the corpus is then represented as a TF-IDF vector for the following cosine similarity computation. In line with previous conceptual IA techniques~\cite{wang2018integrated, gethers:icse12}, LSI further employs singular value decomposition (SVD) on the TF-IDF matrix consisting of TF-IDF representations of all methods in the corpus, and the cosine similarity is computed based on the new dimension-reduced method representations. As for doc2vec, we first train the model utilizing the \textit{distributed memory} algorithm on the CodeSearchNet Java split dataset by concatenating comment tokens with code tokens to maintain consistency with the Transformer-based models (\eg GraphCodeBERT) training process. The doc2vec model can then generate paragraph-based method representations for the constructed IA tasks.

\subsection{\toolname Configurations}
\label{subsec:configs}
By using our approach \toolname, we integrate global dependence information into local code semantics to improve IA, and we set $w = 0.5$ for information balancing. We use GraphCodeBERT as the encoder for the final version of \toolname reported in \ref{rq:athena}, given that it achieves the best IA performance. We also validate the effectiveness of the initial method representations (without embedding propagation) obtained by GraphCodeBERT for conceptual IA (\toolct), and conduct experiments using either call (\toolcd) or class member dependencies (\toolcmd) with GraphCodeBERT in order to quantitatively show the contribution of each type of dependence from the dependence graphs.

We also experimented with different encoders (\ie CodeBERT~\cite{feng2020codebert} and UniXcoder~\cite{guo2022unixcoder}) that are also fine-tuned on the code search task following a similar procedure to the one described in \Cref{sec:approach}, in order to demonstrate the effectiveness of our approach when using other Transformer-based code models. Moreover, we try neighbors of different orders/distances (1--3) when propagating the embeddings based on structural dependence graphs. Additionally, we conduct experiments based on different initial method representations obtained by GraphCodeBERT, including with/without comments, using the output of the [CLS] token to represent methods, and using the pre-trained GraphCodeBERT directly without fine-tuning it on code search. Last, we also fine-tune the pre-trained GraphCodeBERT on the BigCloneBench dataset~\cite{lu2021codexglue} constructed for the clone detection task, following the same procedure provided by Guo \etal~\cite{guo2020graphcodebert}, and employ this fine-tuned GraphCodeBERT to directly generate the probability of whether two methods are semantically similar for the IA task.

\section{Evaluation Results \& Discussion}\label{sec:results}

\subsection{\ref{rq:athena}: \toolname Performance on IA}\label{sec:results_1}

\begin{table}[t]
\vspace{-0.3cm}
\footnotesize
\begin{adjustbox}{width=0.98\linewidth,center}
\begin{minipage}{.5\linewidth}
      \caption{Effectiveness of the baseline techniques (\%).}
      \label{tab:baselineresults}
\begin{tabular}{c|c|ccc|}
\toprule
\textbf{Baseline}                         & \textbf{Settings} & \textbf{mRR}   & \textbf{mAP}   & \textbf{Hit@10} \\ \hline
\multirow{3}{*}{\textbf{TF-IDF}}          & whole             & 49.57          & 25.38          & 70.35          \\
                                          & inner             & 73.86          & 64.69          & 94.61            \\
                                          & outer             & 34.50          & 16.50          & 49.35           \\ \hline
\multirow{3}{*}{\textbf{LSI}}             & whole             & 49.98          & 25.64          & 69.80           \\
                                          & inner             & 74.11          & 64.97          & 94.53   \\
                                          & outer             & 34.85          & 16.68          & 49.45    \\ \hline
\multirow{3}{*}{\textbf{doc2vec}}         & whole             & 43.62 & 19.97 & 58.59            \\
                                          & inner             & 68.93 & 59.05 & 90.97           \\
                                          & outer             & 29.63 & 12.35 & 40.25          \\ \hline
\multirow{3}{*}{\textbf{LSI (+comm)}} & whole             & 50.28          & 26.16          & 70.94     \\
                                          & inner             & 73.83 & 64.69 & 94.61       \\
                                          & outer             & 34.60 & 19.93 & 49.91      \\ \bottomrule
\end{tabular}
\end{minipage}
\begin{minipage}{.5\linewidth}
      \caption{Effectiveness of \toolname (\%).}
      \label{tab:athenaresults}
\begin{tabular}{|c|c|ccc}
\toprule
 \textbf{\toolname Config}                           & \textbf{Settings} & \textbf{mRR}   & \textbf{mAP}   & \textbf{Hit@10} \\ \hline
 \multirow{3}{*}{\textbf{\toolct}}        & whole             & 52.38          & 28.86          & 73.87           \\
                                              & inner             & \textbf{75.94} & \textbf{66.24} & 95.44           \\
                                             & outer             & 40.39          & 21.43          & 58.19           \\ \hline

 \multirow{3}{*}{\textbf{\toolcd}}     & whole             & 54.26          & 30.43          & 76.96           \\
                                                                                & inner             & 75.05          & 65.52          & 95.38           \\
                                                                                    & outer             & 42.50          & 22.70          & 60.95           \\ \hline
         \multirow{3}{*}{\textbf{\toolcmd}}    & whole             & 59.55          & 34.50          & 80.50           \\
                                                                                 & inner             & 75.91          & 66.22          & 95.32           \\
                                                                                 & outer             & 42.93          & 22.02          & 59.92           \\  \hline
                                                                                    \multirow{3}{*}{\textbf{\toolname}} & whole             & \textbf{60.32} & \textbf{35.19} & \textbf{81.48}  \\
                                                                               & inner             & 75.59          & 65.94          & \textbf{95.80}  \\
                                                                                 & outer             & \textbf{45.07} & \textbf{23.41} & \textbf{61.59}  \\ \bottomrule
\end{tabular}
\end{minipage}
\end{adjustbox}
\vspace{-0.3cm}
\end{table}

\Cref{tab:athenaresults} presents \toolnames performance (\%) on our \bmname benchmark whereas \Cref{tab:baselineresults} reports results for three baseline models for IA. All of these models take code-only information (\ie without comments) as input except LSI (+comm.), and we will show the performance of \toolname (+comm.) in the~\ref{rq:ablation} ablation study. The results in~\Cref{tab:baselineresults} reveal that the LSI model achieves the highest effectiveness among the baseline models across all three settings. Given the effect of the number of related topics on LSI's performance, we experimented with varying numbers of related topics (from 0 to 2,000 in increments of 100) and selected the one with the best performance (1,300) for the final LSI configuration. Moreover, LSI only slightly outperforms TF-IDF on the three metrics, indicating that the advantage is not significant if high-level code semantics are extracted through SVD. Surprisingly, the doc2vec model performs worse than LSI. This could be due to the fact that the IR-based approaches can directly build corpora and measure the importance of code tokens on the evaluation dataset, and thus excel at keyword matching in favor of IA. In contrast, the deep learning-based model doc2vec is primarily trained for high-level semantic understanding rather than keyword matching, with the evaluation set unknown, and it struggles with understanding code intent compared to Transformer-based code models.
In addition, we add comment information to the input for the best-performing baseline LSI, but the with-comment version only performs slightly better than the one without comments in Setting 1 (whole), and not in Setting 2 (inner) or Setting 3 (outer) on mRR and mAP, which does not amount to a real improvement for IA. We provide a detailed explanation of this in~\ref{rq:dependence}.

As can be seen from~\Cref{tab:athenaresults}, both \toolct (without embedding propagation) and \toolname outperform LSI with statistical significance (Wilcoxon's paired test, $p<0.05$) on all three metrics across all settings, and their improvements in Setting 1 (whole) can mainly be attributed to the improvements in Setting 3 (outer). Specifically, \toolct improves over LSI by 2.40\%/3.22\% mRR/mAP in Setting 1, and 5.54\%/4.75\% mRR/mAP in Setting 3. In fact, LSI performs quite well in Setting 2 (inner) because of its proficiency in keyword matching and the observation that keyword overlap is more common among methods within the same class as the query. Yet the Transformer-based model GraphCodeBERT excels in understanding the underlying code semantics, resulting in superior performance of \toolct in both Setting 2 and Setting 3. However, the improvements from Settings 2 and 3 do not all contribute to the performance gain for Setting 1. The reason behind this is that LSI tends to rank all methods in the same class as the query higher than those in other classes, and methods in the same class are more likely to be actually affected, as indicated by the ratio of ground-truth impact set size to corpus size in~\Cref{tab:dataset}. Consequently, LSI achieves better relative performance in Setting 1 (\ie a smaller improvement margin for \toolct) than in Setting 3, but this does not change the relative positions of methods within the same class (Setting 2) or methods in different classes (Setting 3). More evidence supporting this explanation is provided in~\ref{rq:dependence}. In addition, when integrating global dependence information into local code semantics, \toolname substantially outperforms LSI by 10.34\%/9.55\% and 10.22\%/6.73\% mRR/mAP in Settings 1 and 3, respectively. \toolname considers neighbor methods within two orders (hops) in dependence graphs for embedding propagation.

\subsection{\ref{rq:dependence}: The Impact of Call Dependence and Class Member Dependence}

In~\Cref{tab:athenaresults}, we also present the performance of \toolname when utilizing either the call (\ie \toolcd) or the class member dependencies (\ie \toolcmd) for embedding propagation based on dependence graphs, which allows us to investigate how each type of dependency contributes to the effectiveness of \toolname in IA. By comparing both \toolcd and \toolcmd with \toolct, we observed that both of them outperform \toolct, and their improvements in Setting 1 (whole) are also attributable to the improvements in Setting 3 (outer). This confirms the accuracy of our dependence graph generator when capturing either the call or the class member dependence.

Although \toolcd and \toolcmd obtain comparable results in Setting 2 and in Setting 3, in Setting 1 \toolcmd outperforms \toolcd by 5.29\%/4.07\% on mRR/mAP. This is because in \toolcmd, the query method is integrated with the information from all the other methods in the same class. As such, it ranks all these methods higher than those in other classes, as previously described in~\Cref{sec:results_1}. To further support this explanation, we experimented with another strategy for considering only class member dependence. Instead of using embedding propagation, we directly reduce the cosine distance between the query method and each method within the same class as the query by 50\% for IA. The results are quite good in Setting 1 (60.72\%/37.23\% mRR/mAP), but as expected it behaves exactly the same as \toolct in Settings 2 and 3, because while all methods in the same class are drawn closer to the query, the relative positions of methods in the same class or those in other classes remain unchanged. In addition, when comparing both \toolcd and \toolcmd with \toolname, both contribute to \toolnames effectiveness, particularly in Settings 1 and 3.

\subsection{\ref{rq:ablation}: Ablation Study}

\begin{table}[t]
\caption{Ablation study of \toolname on mRR and mAP (\%).}
\setlength{\tabcolsep}{2.8pt}
\label{tab:ablation_results}
\begin{adjustbox}{width=\linewidth,center}
\begin{tabular}{c|cccc|cccc|cc|cc|cc|cc}
\toprule
\multirow{3}{*}{\textbf{Settings}} & \multicolumn{4}{c|}{\textbf{Encoders}}                                                              & \multicolumn{4}{c|}{\textbf{\# neighbor orders}}                                                       & \multicolumn{2}{l|}{\multirow{2}{*}{\textbf{{[}CLS{]} token}}} & \multicolumn{2}{c|}{\multirow{2}{*}{\textbf{pretrain-only}}} & \multicolumn{2}{l|}{\multirow{2}{*}{\textbf{+comm.}}} & \multicolumn{2}{l}{\multirow{2}{*}{\textbf{clone detect.}}} \\ \cline{2-9}
                          & \multicolumn{2}{c|}{\textbf{CodeBERT}}           & \multicolumn{2}{c|}{\textbf{UniXcoder}} & \multicolumn{2}{c|}{\textbf{1 order}}            & \multicolumn{2}{c|}{\textbf{3 orders}} & \multicolumn{2}{l|}{}                                 & \multicolumn{2}{c|}{}                          & \multicolumn{2}{l|}{}                           & \multicolumn{2}{l}{}                            \\ \cline{2-17}
                          & \textbf{mRR} & \multicolumn{1}{c|}{\textbf{mAP}} & \textbf{mRR}       & \textbf{mAP}       & \textbf{mRR} & \multicolumn{1}{c|}{\textbf{mAP}} & \textbf{mRR}       & \textbf{mAP}      & \textbf{mRR}              & \textbf{mAP}              & \textbf{mRR}           & \textbf{mAP}          & \textbf{mRR}           & \textbf{mAP}           & \textbf{mRR}           & \textbf{mAP}           \\ \hline
whole                     & 58.40        & \multicolumn{1}{c|}{33.37}        & 60.19              & 34.61              & 59.42        & \multicolumn{1}{c|}{34.33}        & 59.90              & 34.73             & 56.36                     & 32.10                     & 59.92                  & 32.86                 & 59.92                  & 34.96                  & 47.26                  & 22.72                  \\
inner                     & 74.68        & \multicolumn{1}{c|}{64.74}        & 75.87              & 66.18              & 75.95        & \multicolumn{1}{c|}{66.26}        & 74.94              & 65.20             & 73.74                     & 63.83                     & 75.62                  & 65.94                 & 75.12                  & 65.37                  & 71.18                  & 61.11                  \\
outer                     & 43.09        & \multicolumn{1}{c|}{22.08}        & 43.93              & 22.64              & 43.80        & \multicolumn{1}{c|}{22.56}        & 44.66              & 23.12             & 42.67                     & 22.12                     & 41.48                  & 19.99                 & 45.11                  & 23.54                  & 32.42                  & 14.43                  \\ \bottomrule
\end{tabular}
\end{adjustbox}
\vspace{-1em}
\end{table}

\Cref{tab:ablation_results} illustrates the various configurations of \toolname for the ablation study. Specifically, we first conducted experiments using different pre-trained Transformer-based code models, namely CodeBERT and UniXcoder. Both of them were also fine-tuned on the code search task in order to transfer additional knowledge learned from code search to IA, similar to our approach with GraphCodeBERT. Also, we follow the procedures recommended in the corresponding papers for fine-tuning and IA evaluation (\eg the AST is only used for UniXcoder pre-training, but not for fine-tuning and evaluation). Since CodeBERT only considers sequential code information during pre-training and fine-tuning, the method representations obtained by CodeBERT are not as meaningful as those obtained by GraphCodeBERT, which results in poorer performance than \toolname on IA. On the other hand, UniXcoder's IA results are comparable to those of GraphCodeBERT in Setting 1 (whole), but it does not perform as well as GraphCodeBERT in Setting 3 (outer). This may be due to the fact that UniXcoder only utilizes AST information in pre-training, but not in fine-tuning and evaluation, unlike GraphCodeBERT, which utilizes data flow in all these phases, thus benefiting the understanding of the underlying code intent. Moreover, we experimented with neighbor methods of different orders (1 and 3) for embedding propagation for IA, and the results showed that utilizing neighbor methods within two orders (\toolname) is the optimal choice. Although considering the third order involves more dependent methods and requires more computational resources, it does not improve the IA performance.

Moreover, instead of taking the average output of all hidden states from the final layer, we experimented with using the output of the [CLS] token of the Transformer-based model (\ie GraphCodeBERT) as the initial method representation for \toolname. While the output of the [CLS] token is widely used for code understanding-related tasks (\eg code search), taking the average output of all hidden states is more suitable for representing code semantics for IA, according to the results shown in~\Cref{tab:athenaresults} and~\Cref{tab:ablation_results}. We also conducted experiments by removing the code search fine-tuning of \toolname and using the pre-trained GraphCodeBERT directly for initial method embedding extraction, but the pre-trained GraphCodeBERT is less effective than the fine-tuned one (\toolname) for IA especially in Setting 3 (by 3.59\%/3.42\% mRR/mAP). The reason is that during code search fine-tuning, the code is mapped closer to its corresponding NL description, further enhancing the model's ability to understand the underlying code intent and thereby improving \toolnames effectiveness. In addition, we add comment information to the input of \toolname, but the benefit is not obvious, probably because our IA evaluation benchmark \bmname directly collects developer-written methods from commit histories, resulting in some methods having (documentation) comments while others do not (a realistic setting for IA), which may negatively affect the similarity computation between methods. However, the CodeSearchNet dataset used for code search fine-tuning is well curated to ensure that each code snippet is paired with its corresponding NL description (\ie the first line of the documentation comment). Therefore, for the sake of efficiency, our final version of \toolname takes code-only information as input, with data flow extracted, for IA.

\begin{table}[t]
\scriptsize
\caption{Evaluation results (\%) of LSI, \toolct, and \toolname on the filtered \bmname benchmark and its tangled counterpart.}
\vspace{-0.3cm}
\label{tab:benchmark}
\begin{adjustbox}{width=0.97\linewidth,center}
\begin{tabular}{l|lccc|lccc|lccc}
\toprule
\multirow{3}{*}{\textbf{Settings}} & \multicolumn{4}{c|}{\textbf{LSI}}                                                                                                    & \multicolumn{4}{c|}{\textbf{\toolct}}                                                                                                 & \multicolumn{4}{c}{\textbf{\toolname}}                                                                                               \\ \cline{2-13}
                                   & \multicolumn{2}{c|}{\textbf{tangled}}                         & \multicolumn{2}{c|}{\textbf{untangled}}                              & \multicolumn{2}{c|}{\textbf{tangled}}                         & \multicolumn{2}{c|}{\textbf{untangled}}                              & \multicolumn{2}{c|}{\textbf{tangled}}                         & \multicolumn{2}{c}{\textbf{untangled}}                              \\ \cline{2-13}
                                   & \textbf{mRR}              & \multicolumn{1}{l|}{\textbf{mAP}} & \multicolumn{1}{l}{\textbf{mRR}} & \multicolumn{1}{l|}{\textbf{mAP}} & \textbf{mRR}              & \multicolumn{1}{l|}{\textbf{mAP}} & \multicolumn{1}{l}{\textbf{mRR}} & \multicolumn{1}{l|}{\textbf{mAP}} & \textbf{mRR}              & \multicolumn{1}{l|}{\textbf{mAP}} & \multicolumn{1}{l}{\textbf{mRR}} & \multicolumn{1}{l}{\textbf{mAP}} \\ \hline
\multicolumn{1}{c|}{whole}         & \multicolumn{1}{c}{52.94} & \multicolumn{1}{c|}{17.51}        & \textbf{58.42}                   & \textbf{18.88}                    & \multicolumn{1}{c}{54.93} & \multicolumn{1}{c|}{19.56}        & \textbf{60.55}                   & \textbf{20.86}                    & \multicolumn{1}{c}{64.56} & \multicolumn{1}{c|}{23.71}        & \textbf{68.36}                   & \textbf{24.88}                   \\
\multicolumn{1}{c|}{inner}         & \multicolumn{1}{c}{80.72} & \multicolumn{1}{c|}{70.36}        & \textbf{82.17}                   & \textbf{71.50}                    & \multicolumn{1}{c}{81.37} & \multicolumn{1}{c|}{69.03}        & \textbf{82.67}                   & \textbf{70.69}                    & \multicolumn{1}{c}{81.81} & \multicolumn{1}{c|}{69.16}        & \textbf{83.65}                   & \textbf{71.05}                   \\
\multicolumn{1}{c|}{outer}         & \multicolumn{1}{c}{37.72} & \multicolumn{1}{c|}{11.45}        & \textbf{42.89}                   & \textbf{12.80}                    & \multicolumn{1}{c}{41.31} & \multicolumn{1}{c|}{15.10}        & \textbf{46.06}                   & \textbf{15.72}                    & \multicolumn{1}{c}{47.53} & \multicolumn{1}{c|}{16.41}        & \textbf{50.65}                   & \textbf{17.09}                   \\ \bottomrule
\end{tabular}
\end{adjustbox}
\vspace{-2em}
\end{table}

In addition, we replace code search with clone detection to use it as a proxy for IA. Specifically, we fine-tuned GraphCodeBERT for clone detection following the same pipeline recommended by Guo \etal~\cite{guo2020graphcodebert}. Instead of generating separate code embeddings, the model directly produces the probability of whether two code snippets can yield similar results, and as a result, the embedding propagation strategy cannot be applied. Therefore, we utilize the generated probability scores to obtain a ranked list for IA and compare it with \toolct (without embedding propagation). However, from~\Cref{tab:athenaresults} and~\Cref{tab:ablation_results}, we observe that using clone detection as a proxy is less effective than \toolct using code search.

\subsection{\ref{rq:tangled}: \toolname and Baseline Performance on the Tangled Benchmark}

In~\Cref{tab:benchmark}, we present the evaluation results of the best-performing baseline LSI, \toolct, and \toolname on the filtered \bmname and its corresponding tangled counterpart using the mRR and mAP metrics. Specifically, after comparing our IA benchmark \bmname with its tangled counterpart, we extract the tasks with inconsistent (query, ground-truth impact set) pairs and conduct experiments on these filtered tasks from \bmname (untangled) and its tangled counterpart, respectively. The statistics of the filtered datasets are described in~\Cref{sec:benchmark}. As observed in~\Cref{tab:benchmark}, there is a substantial performance difference between untangled \bmname and its tangled counterpart across all three settings when using any of the models, especially on mRR (ranging from 3.80\% to 5.62\% in Setting 1). However, existing IA benchmarks are typically built from tangled/original commits, which affects the reliability of the evaluation results of previous IA techniques. Moreover, as expected, each of the three models performs better on untangled \bmname than on the tangled version across all three settings. The reason is that each co-changed set in \bmname was manually verified to address one single concern, ensuring that the methods within it are truly impacted by each other. In contrast, the tangled counterpart is built from original/unvetted commits, and the methods within each co-changed set may not all contribute to one concern, and thus are not necessarily impacted by each other. Therefore, identifying the methods that are truly impacted with respect to the query is harder for each of the representative models.

\subsection{\ref{rq:impact}: Qualitative Analyses on IA Tasks}

\begin{table}[t]
\centering
\vspace{-1em}
\scriptsize
\caption{Effectiveness (\%) of \toolname and the LSI baseline for each software system in Setting 1 (whole).}
\label{tabs:repo_results}
\begin{tabular}{l|c|c|c|ccc|ccc}
\toprule
\multicolumn{1}{c|}{\multirow{2}{*}{\textbf{Repo Name}}} & \multirow{2}{*}{\textbf{LOC(k)}} & \multirow{2}{*}{\textbf{\# Commits}} & \multirow{2}{*}{\textbf{\# queries}} & \multicolumn{3}{c|}{\textbf{\toolname}}              & \multicolumn{3}{c}{\textbf{LSI}}                   \\ \cline{5-10}
\multicolumn{1}{c|}{}                                    &                                  &                                      &                                      & \textbf{mRR}   & \textbf{mAP}   & \textbf{HIT@10} & \textbf{mRR}    & \textbf{mAP}   & \textbf{HIT@10} \\ \hline
ant-ivy                                                  & 412.3                            & 176                                  & 785                                  & \textbf{50.19} & \textbf{26.47} & \textbf{72.36}  & 39.79           & 18.48          & 60.64           \\
archiva                                                  & 361.2                            & 2                                    & 43                                   & \textbf{70.81} & \textbf{32.17} & \textbf{88.37}  & 69.39           & 10.17          & \textbf{88.37}  \\
commons-bcel                                             & 168.3                            & 18                                   & 138                                  & \textbf{66.07} & \textbf{30.79} & \textbf{87.68}  & 57.76           & 21.68          & 71.74           \\
commons-beanutils                                        & 67.5                             & 11                                   & 42                                   & 65.64          & \textbf{44.58} & \textbf{95.24}  & \textbf{67.67}  & 43.64          & 83.33           \\
commons-codec                                            & 55.1                             & 8                                    & 41                                   & \textbf{67.78} & \textbf{52.79} & \textbf{90.24}  & 57.65           & 34.91          & 78.05           \\
commons-collections                                      & 136.3                            & 15                                   & 73                                   & \textbf{47.84} & \textbf{24.80} & \textbf{84.93}  & 41.43           & 18.85          & 68.49           \\
commons-compress                                         & 147.3                            & 61                                   & 260                                  & \textbf{51.67} & \textbf{32.99} & \textbf{68.85}  & 45.26           & 23.73          & 66.15           \\
commons-configuration                                    & 72.9                             & 65                                   & 253                                  & \textbf{56.89} & \textbf{36.87} & \textbf{78.26}  & 41.04           & 24.75          & 58.10           \\
commons-dbcp                                             & 55.6                             & 21                                   & 91                                   & \textbf{67.17} & \textbf{52.55} & \textbf{92.31}  & 61.73           & 46.65          & 84.62           \\
commons-digester                                         & 89.7                             & 8                                    & 22                                   & \textbf{38.65} & \textbf{29.07} & \textbf{77.27}  & 28.05           & 23.86          & 45.46           \\
commons-io                                               & 102.5                            & 19                                   & 58                                   & \textbf{64.34} & \textbf{49.13} & \textbf{91.38}  & 52.16           & 32.53          & 75.86           \\
commons-jcs                                              & 164                              & 26                                   & 221                                  & \textbf{70.35} & \textbf{26.10} & \textbf{85.07}  & 61.02           & 18.86          & 76.92           \\
commons-lang                                             & 192.5                            & 36                                   & 115                                  & \textbf{67.16} & \textbf{56.23} & \textbf{89.57}  & 58.38           & 46.66          & 80.87           \\
commons-math                                             & 431.1                            & 124                                  & 589                                  & \textbf{65.93} & \textbf{42.02} & \textbf{87.44}  & 52.43           & 29.20          & 73.35           \\
commons-net                                              & 58.2                             & 44                                   & 171                                  & \textbf{66.59} & \textbf{44.59} & \textbf{84.80}  & 51.02           & 26.35          & 70.18           \\
commons-scxml                                            & 43.8                             & 28                                   & 114                                  & \textbf{50.32} & \textbf{34.62} & \textbf{75.44}  & 45.82           & 31.19          & 72.81           \\
commons-validator                                        & 42.3                             & 12                                   & 35                                   & \textbf{62.74} & \textbf{56.51} & \textbf{85.71}  & 51.70           & 40.29          & 74.29           \\
commons-vfs                                              & 91.2                             & 40                                   & 166                                  & \textbf{55.02} & \textbf{36.62} & \textbf{83.13}  & 51.30           & 35.71          & 74.10           \\
deltaspike                                               & 174.2                            & 2                                    & 5                                    & \textbf{60.98} & \textbf{57.65} & \textbf{60.00}  & 35.04           & 27.25          & \textbf{60.00}  \\
giraph                                                   & 200.6                            & 68                                   & 527                                  & \textbf{70.80} & \textbf{38.40} & \textbf{89.75}  & 59.01           & 26.78          & 81.59           \\
gora                                                     & 132.4                            & 40                                   & 174                                  & \textbf{49.31} & \textbf{26.93} & \textbf{68.97}  & 41.91           & 23.59          & 62.64           \\
jspwiki                                                  & 439.4                            & 1                                    & 12                                   & 87.50          & 40.03          & \textbf{100.00} & \textbf{100.00} & \textbf{70.28} & \textbf{100.00} \\
opennlp                                                  & 293.5                            & 33                                   & 141                                  & \textbf{64.61} & \textbf{40.16} & \textbf{78.72}  & 52.13           & 28.94          & 69.50           \\
parquet                                                  & 177.6                            & 50                                   & 324                                  & \textbf{60.09} & \textbf{25.92} & \textbf{81.48}  & 48.27           & 17.65          & 67.28           \\
systemml                                                 & 4000                             & 2                                    & 5                                    & \textbf{47.15} & \textbf{41.60} & \textbf{80.00}  & 41.63           & 31.25          & 40.00           \\ \bottomrule
\end{tabular}
\vspace{-0.3cm}
\end{table}

We begin our analysis of IA tasks by looking at the performance of our studied techniques across the different studied software projects. \Cref{tabs:repo_results} provides a finer-grained picture of the per-repository improvements our \toolname model achieves over the LSI baseline. As shown, \toolname improves performance on 24 of 25 repositories in terms of mAP and 23 of 25 in terms of mRR in Setting 1 (whole). For the failing repository \textit{commons-beanutils}, we found that \toolname substantially outperforms LSI in Setting 3 (34.97\%/30.58\% vs. 18.68\%/12.37\% mRR/mAP), but not in Setting 2 (75.35\%/64.92\% vs. 88.37\%/80.20\% mRR/mAP). As for the repository \textit{jspwiki}, it contains a single commit with 12 methods in the constructed co-changed set, which corresponds to 12 IA tasks. Among these 12 methods, six methods belong to one class, and the remainder are from another class. After investigating the failed tasks, we found that LSI was able to identify the affected methods quite well when the query and the ground-truth methods had similar code lengths and a lot of keyword overlap, especially when they belonged to the same class. Now that we have examined the performance of \toolname across IA tasks at a repository level, we discuss some exemplars from our benchmark that showcase how incorporating both structural information and semantic information can benefit IA.

\begin{figure*}[t]
	\centering
	\includegraphics[width=0.78\textwidth]{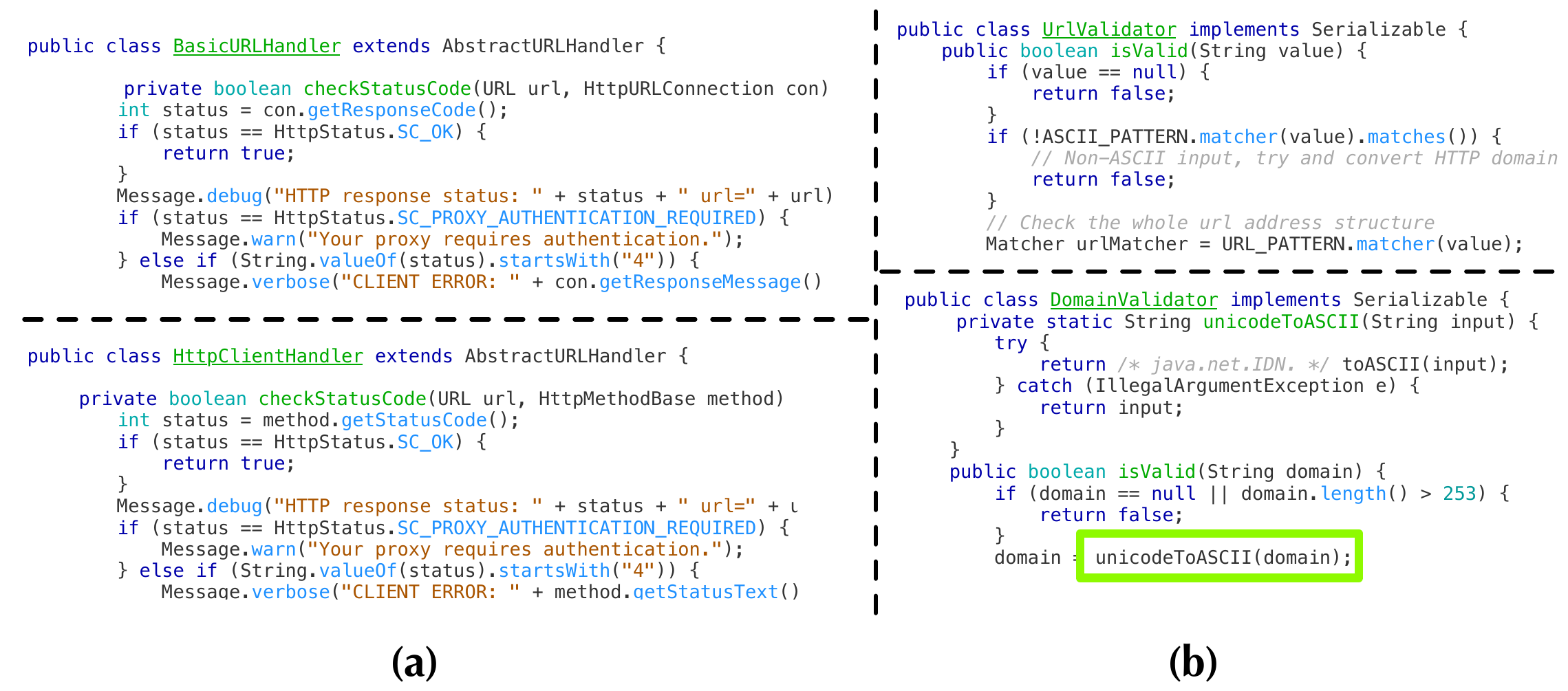}
	\caption{Two qualitative examples illustrating the effectiveness of \toolname. Left: a query method and an impacted method that are conceptually coupled but structurally independent. Right: a query method whose impact set is recovered through embedding propagation over call and class member dependencies.}
	\Description{Two side-by-side Java code listings. The left listing shows the
	checkStatusCode methods of the BasicURLHandler and HttpClientHandler classes, which
	share almost identical logic but no structural dependence. The right listing shows
	the isValid method of UrlValidator together with the unicodeToASCII and isValid
	methods of DomainValidator, where isValid calls unicodeToASCII.}
	\label{fig:example}
\end{figure*}

\textbf{Example 1: The Importance of Semantics.} The left side of \Cref{fig:example} shows two methods from different classes. The top method \texttt{\small checkStatusCode\_URL\_HttpURLConnection} from class \texttt{\small BasicURLHandler} is the query method, and the bottom method \texttt{\small checkStatusCode\_URL\_HttpMethodBase} is in the corresponding ground-truth impact set. This is representative of conceptual coupling~\cite{poshyvanyk2009using}, where the concepts of the two methods --- \ie both performing a check on a status code --- couple them together, making it more likely that a change in one would result in a change in the other. Utilizing the semantic information between the methods, either through traditional LSI or a Transformer-based neural model, is necessary to determine that these two methods are highly related. Since they are not structurally dependent (via call or class member dependencies), a structural dependence-only approach is likely to fail in this scenario.

\textbf{Example 2: The Importance of Richer Semantics and Integration of Dependence Graphs.} The right side of \Cref{fig:example} illustrates a scenario with three methods from two different classes, where the method \texttt{\small isValid} from the class \texttt{\small UrlValidator} is the query, and the methods \texttt{\small unicodeToASCII} and \texttt{\small isValid} from the class \texttt{\small DomainValidator} are in the ground-truth impact set. In this scenario, the baseline LSI places the \texttt{\small unicodeToASCII} method far down the ranked list, at position 589, due to the limited keyword overlap. When using \toolct (without embedding propagation), which leverages GraphCodeBERT for better code understanding, the position of the \texttt{\small unicodeToASCII} method improves to 137. However, this is still deep in the list, which means developers might need substantial effort to locate this method. Remarkably, our \toolname places it at position 36, significantly outperforming the baseline. To understand why this occurred, we found that the method \texttt{\small isValid} in the \texttt{\small DomainValidator} class calls the \texttt{\small unicodeToASCII} method, which means that these two methods have both call and class member dependencies. Through embedding propagation of \toolname, the \texttt{\small unicodeToASCII} method is updated with information from the \texttt{\small isValid} method (in the \texttt{\small DomainValidator} class) that is more semantically similar to the query. This additional information helps improve the rank of the ground truth, even though there is no direct dependence relationship between the query and \texttt{\small unicodeToASCII}.

As can be observed from these examples, there are clear benefits when code understanding is enhanced by the Transformer-based neural model and structural dependence graphs, and we saw this pattern hold after investigating additional cases where \toolname outperforms the baseline LSI. The contextual information obtained from the global call/class member dependencies among methods enriches the original semantics of the methods, which indeed helps to identify the impact set associated with the given query.
\section{Threats to Validity}\label{sec:threats}

\subsection{Internal Validity}
To reduce potential issues from internal threats to validity, we experimented with three different DL models when validating our proposed approach of incorporating program dependence graph information into local code semantics to improve IA. Additionally, we constructed our benchmark from commits that have been manually annotated and had the changes made to fix bugs untangled from other changes, such as documentation changes, to ensure that our benchmark is more reliable.

\subsection{External Validity}
To lessen the potential for threats to external validity, we used a significantly larger set of projects --- \imrepos, compared to previous work that used around five --- and tested our approach across different DL models to show generalizability. One potential issue with generality is that we only evaluated our approach on Java and Apache projects; therefore, our approach may not generalize to other programming languages such as Python, or to different types of projects. However, the DL models we used have shown success across multiple programming languages, and so most likely the same would apply to our approach.
\section{Conclusion}\label{sec:conclusion}

In this paper, we introduce \toolname, a novel technique for impact analysis that combines Transformer-based neural code semantics with structural dependence graphs. Additionally, we established a large benchmark for impact analysis, \bmname, which is built from manually verified, untangled bug fixing commits. On our new benchmark, \toolname demonstrates significant improvements over the simple conceptual baseline (+10.34\% mRR, +9.55\% mAP, and +11.68\% HIT@10) and exhibits robust performance across software systems, with 23 out of \imrepos systems showing improvement in mRR. Furthermore, our analysis reveals that \toolnames performance boost lies in its ability to more effectively identify impacted methods when they are outside the query method's class.

\begin{acks}
This research has been supported in part by the following NSF grants: CCF-2311469, CCF-2311468, CNS-2132281, CCF-2007246, and CCF-1955853. We also acknowledge support from Cisco Systems. Any opinions, findings, and conclusions expressed herein are the authors' and do not necessarily reflect those of the sponsors.
\end{acks}

\bibliographystyle{ACM-Reference-Format}
\bibliography{main.bib}

\end{document}